\documentclass{PoS}

\usepackage{epsfig,multicol,multirow}
\usepackage{amsmath,subfigure,latexsym,amssymb}

\def\gsi{\gtrsim}
\newcommand{\gsim}{\mathop{\gsi}}

\newcommand{\eps}{\epsilon}
\newcommand{\la}{\langle}
\newcommand{\ra}{\rangle}

\newcommand{\beq}{\begin{equation}}
\newcommand{\eeq}{\end{equation}}

\PoS{PoS(LAT2005)138}

\title{Overlap hypercube fermions in QCD with light quarks}

\ShortTitle{Overlap HF in QCD}

\author{\speaker{Wolfgang Bietenholz}\thanks{{Preprint ~ HU-EP-05/38, BI-TP 2005/29, SFB/CPP-05-39. ~~ $\chi$LF Collaboration}}\\
  Institut f\"{u}r Physik \\ 
  Humboldt Universit\"{a}t zu Berlin \\
  Newtonstr.\ 15, D-12489 Berlin, Germany \\
  E-mail: \email{bietenho@physik.hu-berlin.de}}

\author{Stanislav Shcheredin \\
Fakult\"{a}t f\"{u}r Physik \\ 
Universit\"{a}t Bielefeld \\
D-33615 Bielefeld, Germany \\
E-mail: \email{shchered@physik.hu-berlin.de}}

\abstract{We report on simulation results with overlap hypercube
fermions (overlap HF) --- a type of exactly chiral lattice fermions ---
and their link to chiral perturbation theory.
We first sketch the construction of the overlap HF
and discuss its high level of locality.
Next we show applications in the $p$-regime of QCD, where we evaluate
$m_{\pi}$, $m_{\rho}$, the quark mass according to the PCAC relation, 
the renormalisation constant $Z_{A}$ and the pion decay constant $F_{\pi}$
as functions of the bare quark mass.
$F_{\pi}$ is then reconsidered at even smaller quark masses
in the $\eps$-regime, along with the scalar condensate $\Sigma$.
In that context we also discuss results for the topological charges 
and susceptibility.}

\FullConference{XXIIIrd International Symposium on Lattice Field Theory\\

                 25-30 July 2005\\

                 Trinity College, Dublin, Ireland}

\begin{document}

\section{Overlap Hypercube Fermions}\label{section1}

For free fermions, perfect lattice actions are known analytically
\cite{perfect}. The corresponding lattice Dirac operator can be 
truncated to represent a
free Hypercube Fermion (HF), which still
has excellent scaling and chirality properties \cite{BBCW}.
The HF is gauged by fat links over the shortest lattice paths.
Finally the links are amplified by a factor $u \gsim 1$ to restore criticality
and minimise the violation of the Ginsparg-Wilson relation. 
Due to the truncation and the imperfect gauging procedure, the
scaling behaviour and the chirality are somewhat distorted.
Chirality can be corrected again by inserting the HF
in the overlap formula \cite{Neu} (at lattice spacing $a$)
\beq  \label{overlap}
D_{\rm ov} = \frac{\rho}{a} (1 + A / \sqrt{A^{\dagger}A} ) \ , \quad
A := D_{0} - \frac{\rho}{a} \ , \quad \rho \gsim 1 \ ,
\eeq
where $D_{0}$ is some lattice Dirac operator with 
$D_{0} = \gamma_{5} D_{0}^{\dagger}\gamma_{5}$ \ ($\gamma_{5}$-Hermiticity).

$\bullet$ \ The {\em standard overlap fermion} is obtained by inserting the
Wilson operator, $D_{0} = D_{\rm W}$, which is then drastically changed.
We denote the resulting standard overlap operator as ${\bf D}_{\rm{\bf ov-W}}$.

$\bullet$ \ Here we study the case where the HF is inserted in the overlap
formula (\ref{overlap}), $D_{0} = D_{\rm HF}$. This yields
the operator ${\bf D}_{\rm{\bf ov-HF}}$, which describes the {\em overlap HF}.

In both cases, one arrives at exact solutions to the Ginsparg-Wilson relation,
and therefore at an exact (lattice modified) chiral symmetry \cite{Has-Lusch}.
\footnote{The correctness of the axial anomaly in all 
topological sectors has been verified 
for the standard overlap operator in Ref.\ \cite{DHA}, and for the
overlap HF in Ref.\ \cite{AB}.} 
However, in contrast to $D_{\rm W}$, $D_{\rm HF}$ is approximately chiral
already, hence its transformation by the overlap formula,
$D_{\rm HF} \to D_{\rm ov-HF}$, is only a modest modification.
Therefore, the virtues of the HF are essentially inherited by
the overlap HF \cite{ovHF}. 

Here we are going to show mostly quenched results 
with the standard gauge action at $\beta = 5.85$
(i.e.\ $a \simeq 0.123~{\rm fm}$).
For details of the overlap HF construction --- as well as its locality,
which is superior compared to $D_{\rm ov-W}$ --- we refer to Ref.\
\cite{Stani}. 

\FIGURE{
\vspace*{-2mm}
  \includegraphics[angle=270,width=.49\linewidth]{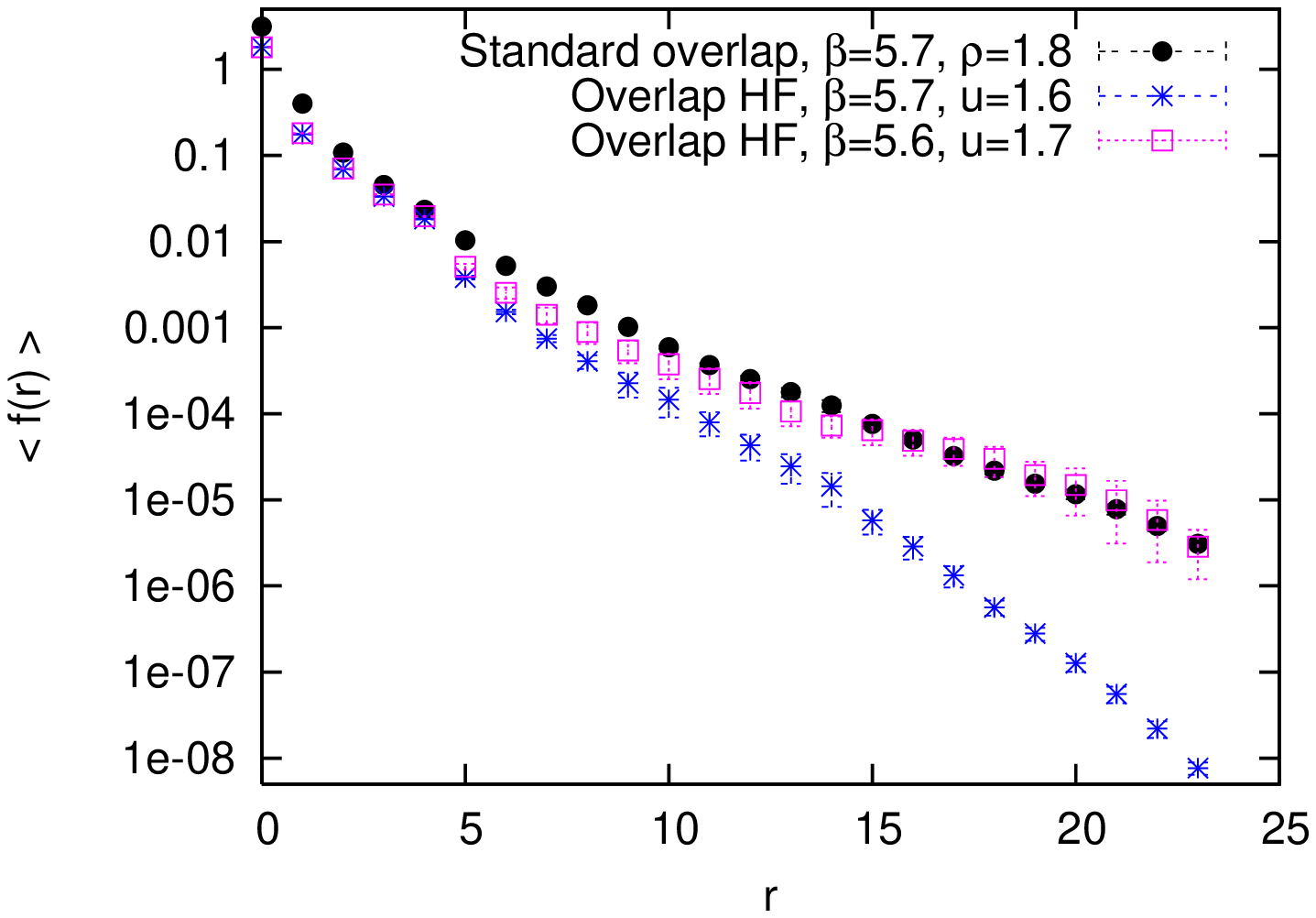}
 \caption{The locality of different overlap fermions, measured by the 
maximal impact of a unit source $\bar \psi_{x}$ on $\psi_{y}$ over a distance
$r = \Vert x -y \Vert_{1}$. At the same value of $\beta$, the overlap HF
is clearly more local, and its locality persists up to $\beta =5.6$,
where the standard overlap fermion collapses.}
\label{localfig}
}

In Fig.\ 1 we illustrate the locality at strong gauge coupling:
at $\beta = 5.7$ 
the standard overlap operator
is still barely local, if one chooses the optimal value $\rho = 1.8$.
Similarly we optimise $u =1.6$ for the HF (at $\rho =1$)
and we still find a clear locality at $\beta = 5.7$, which is in fact stronger
than the one observed for $D_{\rm ov-W}$ at $\beta = 6$ (and
optimal $\rho$) \cite{HJL}. \\
If we proceed to $\beta = 5.6$, the locality collapses for $D_{\rm ov-W}$,
hence in that case the standard overlap formulation does not provide
a valid Dirac operator. On the other hand, if we insert the HF
at $u=1.7$ we still observe locality.
Thus the overlap-HF formulation provides {\em chiral fermions
on coarser lattices.}

\section{Applications in the $p$-Regime}\label{section2}

We first present results in the $p$-regime, which is characterised 
by a box length $L \gg 1/m_{\pi}$, so that the $p$-expansion of chiral 
perturbation theory ($\chi$PT) 
is applicable. 
We consider $\beta=5.85$, a lattice of size $12^{3} \times 24$
and (bare) quark masses of $a m_{q} = 0.01, \ 0.02, \ 0.04,\ 0.06, \
0.08$ and $0.1$.
We computed 100 overlap HF propagators and first evaluated $m_{\pi}$ in three
different ways:
(1) From the pseudoscalar correlator $\la PP \ra$, where 
$P = \bar \psi \gamma_{5} \psi$ 
(2) From the axial correlator
$\la A_{4}A_{4} \ra$, where 
$A_{4} = \bar \psi \gamma_{4} \gamma_{5} \psi$ 
(3) From $\la PP - SS \ra$, where 
$S = \bar \psi \psi$. The subtraction of the scalar density is
useful at small $m_{q}$ to avoid the contamination by zero modes.

\FIGURE{
  \centering
  \includegraphics[angle=270,width=.47\linewidth]{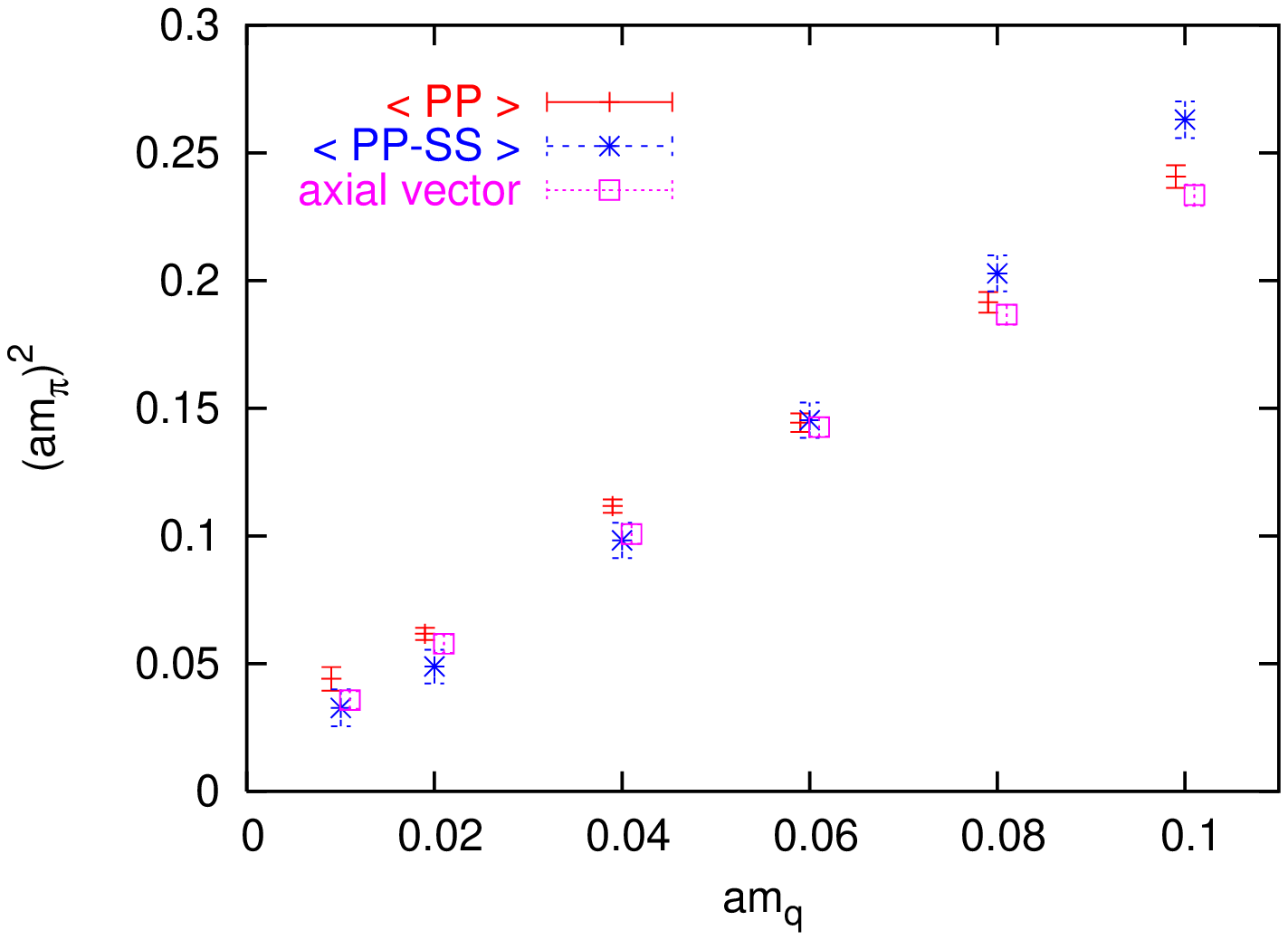}
  \includegraphics[angle=270,width=.47\linewidth]{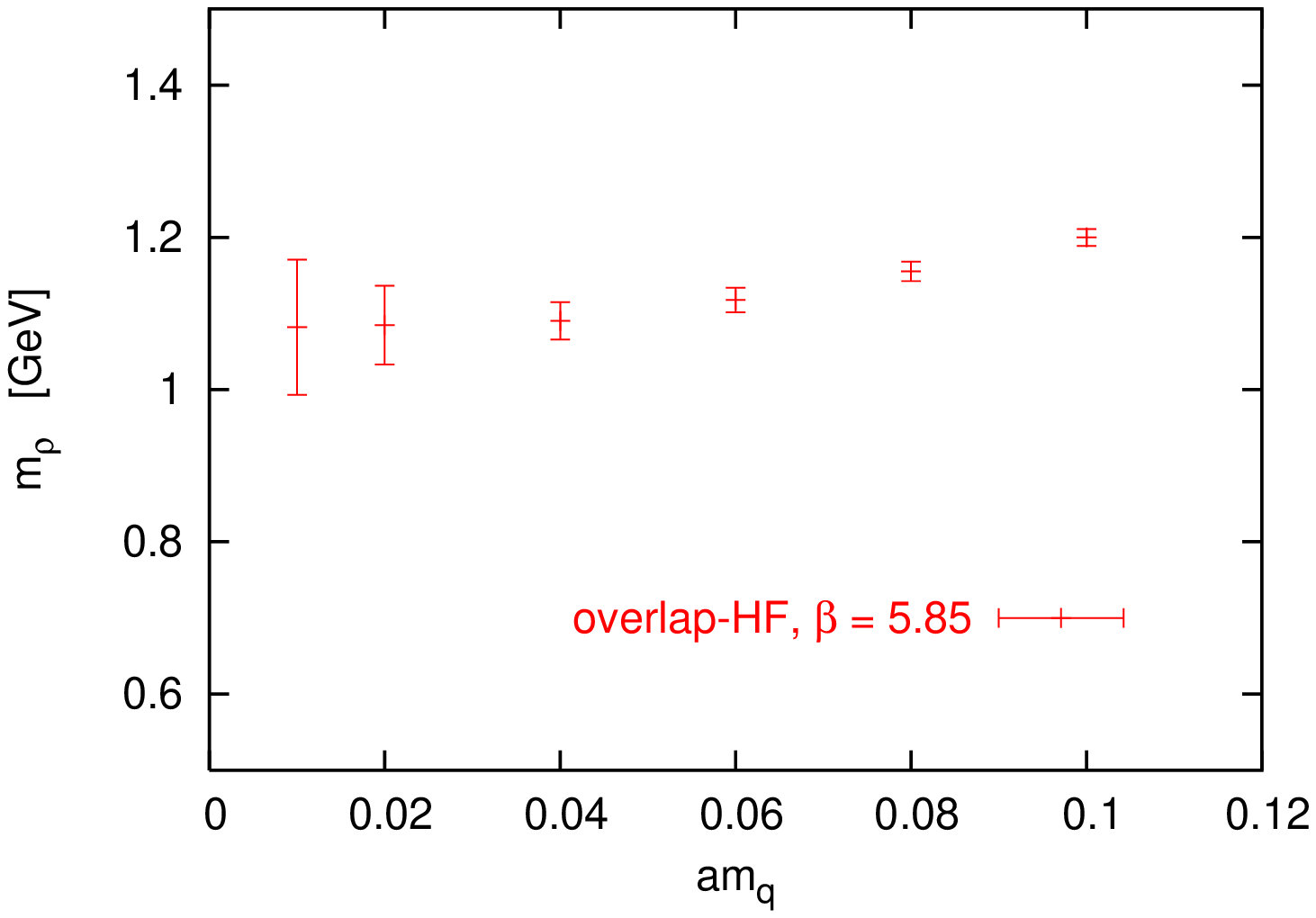}
 \caption{On the left: The pion mass evaluated in various ways.
On the right: the $\rho$-meson mass.}
\label{pirhofig}
}
\noindent
The results are shown in Fig.\ \ref{pirhofig} (left):
they follow well the expected behaviour
$a m_{\pi}^{2} \propto m_{q}$, in particular for
$\la PP - SS \ra$, with a linearly extrapolated intercept
of $a^{2}m_{\pi,PP-SS}^{2}(m_{q}=0) = 0.0001(15)$.
The hierarchy at small $m_{q}$, \ $m_{\pi,PP} > 
m_{\pi,AA} > m_{\pi,PP-SS}$, \ agrees with the literature
\cite{Has02}.
Our smallest pion mass has Compton wave length $\approx L/2$, so
we are at the edge of the $p$-regime.

Fig.\ \ref{pirhofig} (right) shows our results 
for the vector meson mass, with a chiral extrapolation to \\
$m_{\rho} = 1017(39) ~{\rm MeV}$ (quenched results tend to be above
the physical $\rho$ mass).

\FIGURE{
  \centering
  \includegraphics[angle=270,width=.47\linewidth]{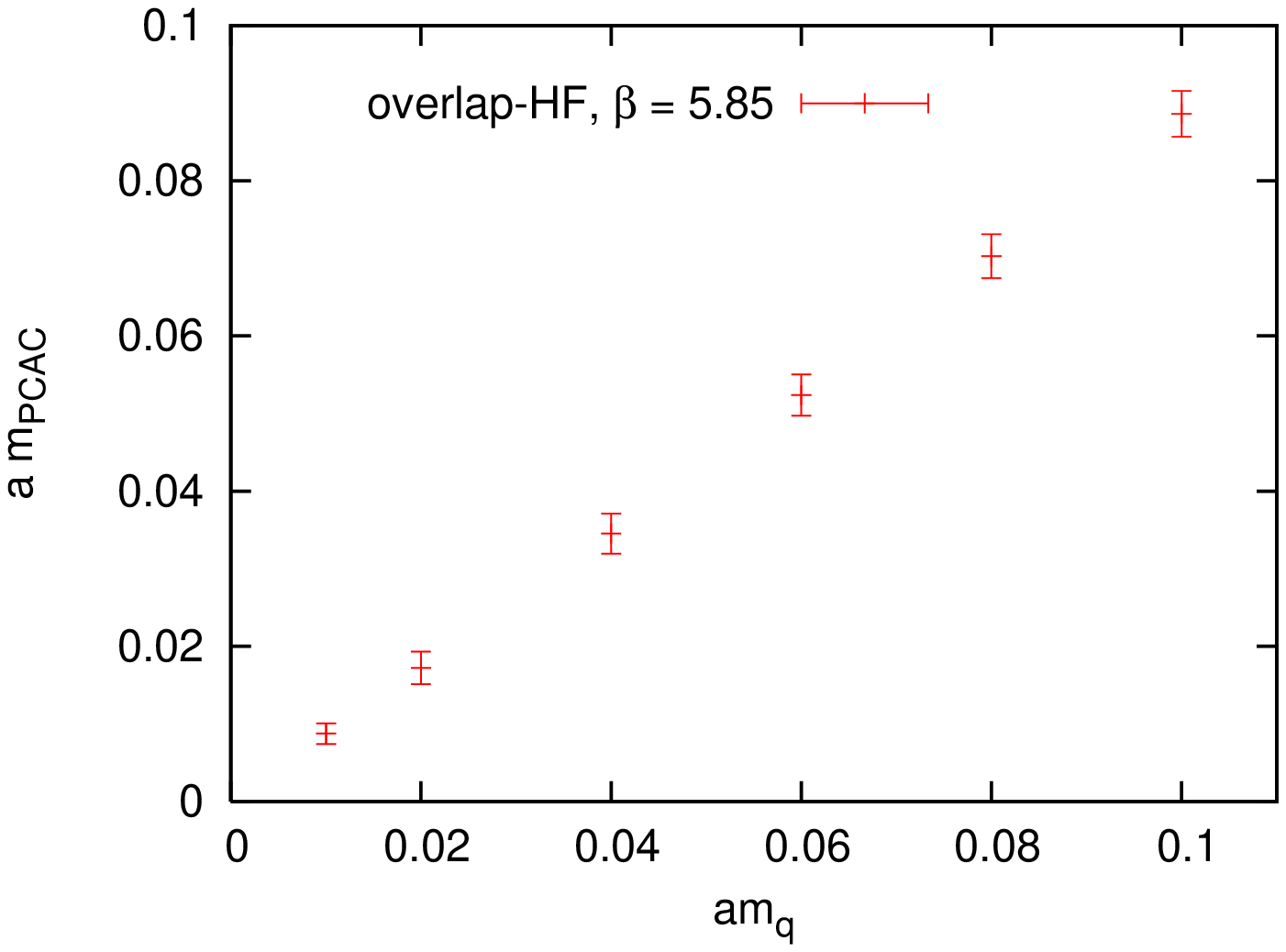}
  \includegraphics[angle=270,width=.47\linewidth]{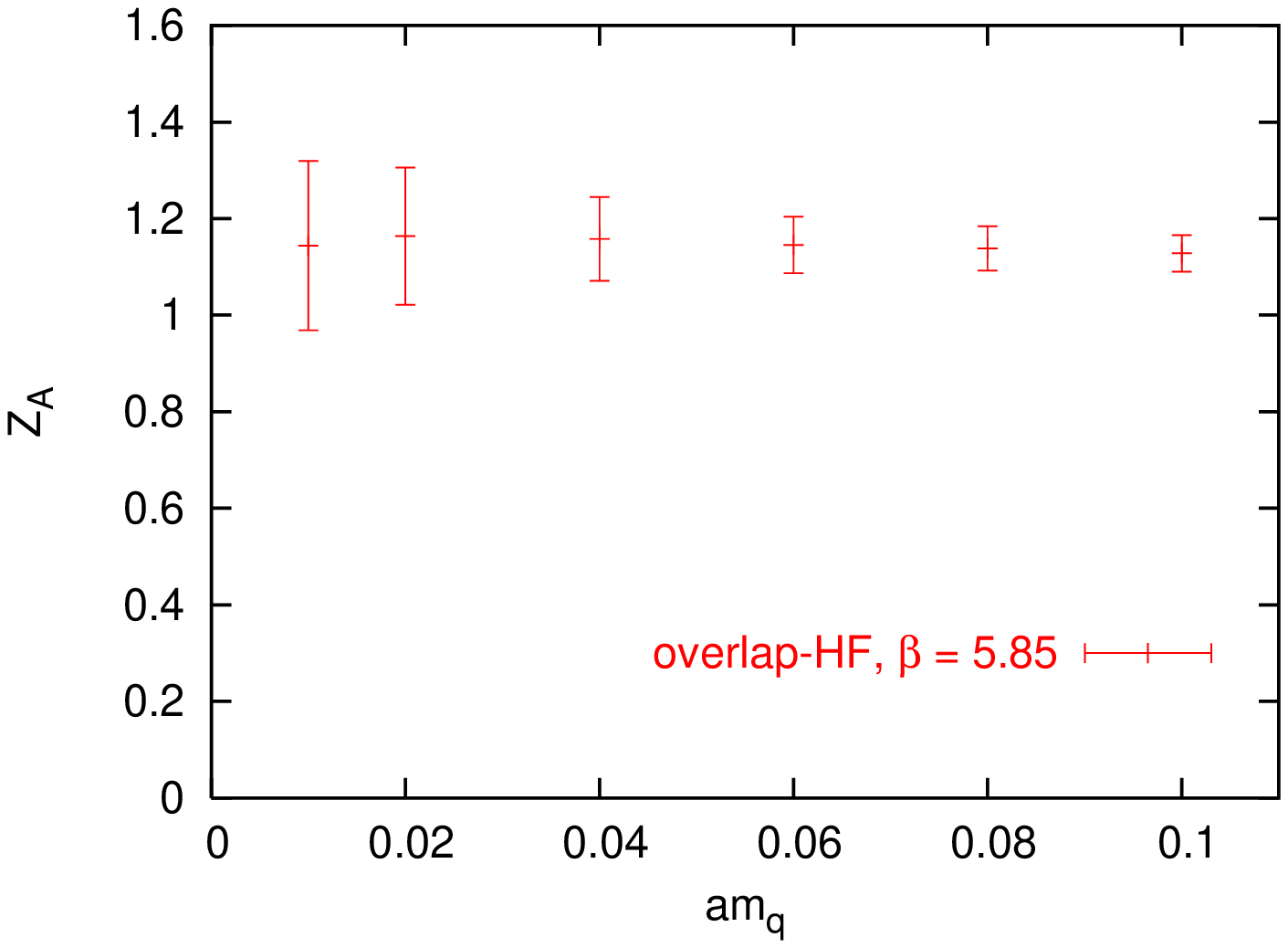}
 \caption{The PCAC quark mass and
the renormalisation constant $Z_{A} = m_{q}/ m_{\rm PCAC}$.}
\label{mPCAC_ZAfig}
}
In Fig.\ \ref{mPCAC_ZAfig} (left)
we consider the quark mass obtained from the 
axial Ward identity,
\beq  \label{AWI}
m_{\rm PCAC} = [ \, \sum_{\vec x} \la \partial_{4} A^{\dagger}(x)
P(0) \ra \, ] \, / \, 2 \, [ \, \sum_{\vec x} \la P^{\dagger} (x) P(0) \ra \, ] \ .
\eeq
\FIGURE{
  \includegraphics[angle=270,width=.46\linewidth]{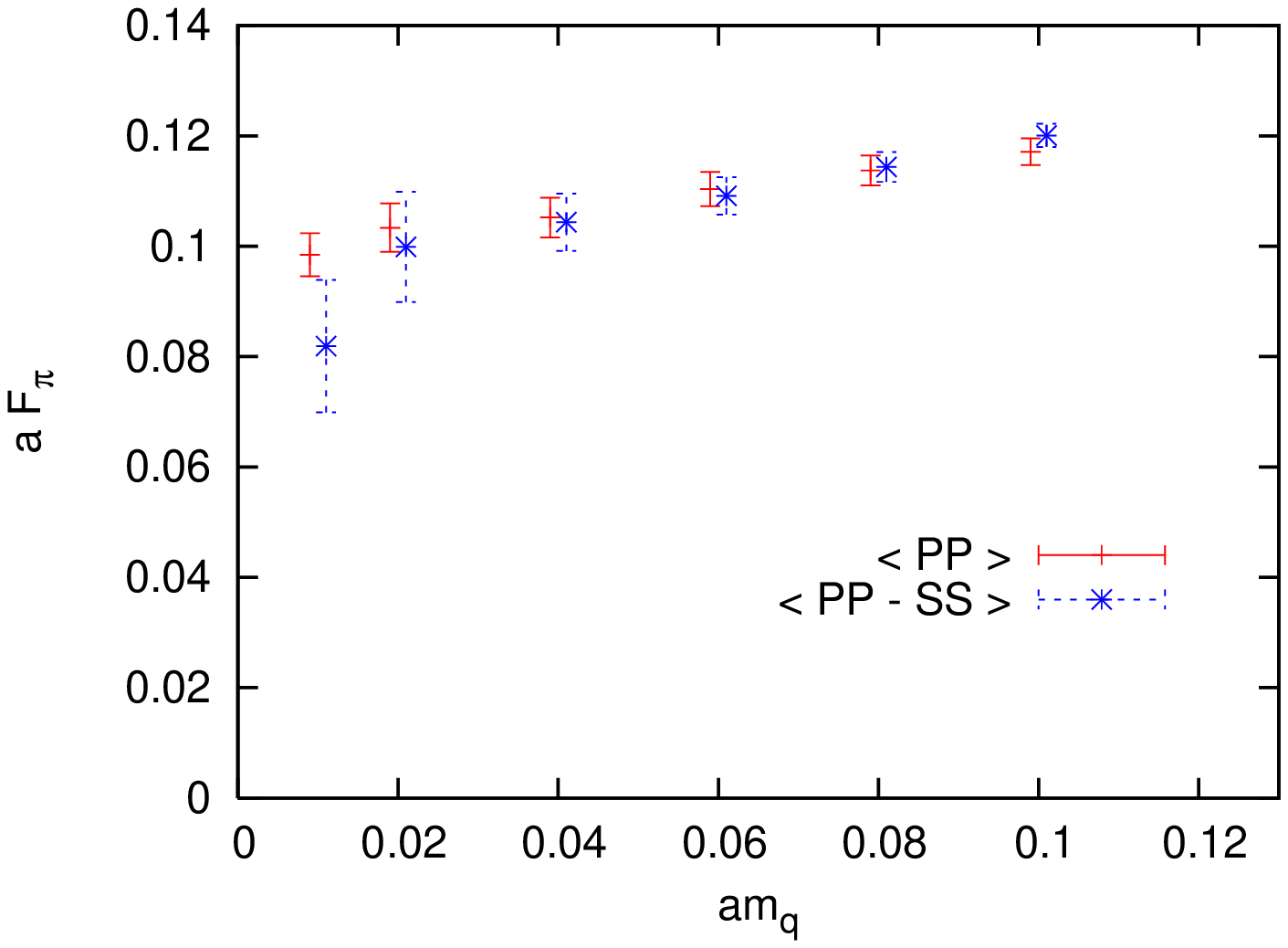}
 \caption{The pion decay constant from a direct evaluation in the
$p$-regime, using the overlap HF.}
\label{Fpifig}
}
\noindent
We observe also here a nearly linear behaviour, with a chiral
extrapolation to $am_{\rm PCAC}(m_{q}=0) = - 0.00029(64)$.
Remarkably, $m_{\rm PCAC}$ is close to $m_{q}$, which is {\em not} 
the case for $D_{\rm ov-W}$ \cite{XLF}. 
Consequently the renormalisation
constant $Z_{A} = m_{q}/ m_{\rm PCAC}$ is close to 1; it
has the chiral extrapolation ${\bf Z_{A} = 1.17(2)}$.
This is in striking contrast to the large $Z_{A}$ factors
obtained for the standard overlap fermions \cite{Berruto,XLF,Jap}.
According to Ref.\ \cite{DHW} the fat link may be helpful for this
favourable feature.

At last we consider \vspace*{1mm}
\ $F_{\pi} = 
\frac{2 m_{q}}{m_{\pi}^{2}} \vert \la 0 \vert P \vert \pi \ra \vert $ \ , \\
by using either $PP$ or $PP-SS$, see Fig.\ \ref{Fpifig}.
The extrapolation to $m_{q}=0$ yields
$F_{\pi , PP} = 111.5 (2.5)$ \\ ${\rm MeV}$, resp.
${\bf{F_{\pi , PP-SS} = 104(9)}} ~ {\rm{\bf{ MeV}}}$, which is
above the physical value 
(taken to the chiral limit) of $86 ~ {\rm MeV}$ \cite{CD}. 
In particular the behaviour of the 
$\la PP-SS \ra$ result at small $m_q$ motivates us to
reconsider $F_{\pi}$ at yet smaller quark masses, which takes
us to the $\eps$-regime.

\section{Applications in the $\eps$-Regime}\label{section3}

In the $\eps$-regime \cite{epsreg} the correlation length exceeds the box length,
$1/m_{\pi} > L$, and
the observables strongly depend on the topological sector.
Our motivation to study this unphysical situation is that it allows
for an evaluation of the Low Energy Constants (LEC) of the chiral
Lagrangian with their values in infinite volume (unfortunately, quenching
brings in logarithmic finite size effects \cite{Dam}).

Random Matrix Theory (RMT) conjectures the densities 
$\rho_{n}^{(\nu )}(z)$ of the lowest
Dirac eigenvalues $\lambda$ in the $\eps$-regime \cite{DamNish},
where $z := \lambda \Sigma V$, $n= 1,2, \dots$ numerates the lowest 
non-zero eigenvalues and $\nu$ is the fermion index, 
which is identified with the topological charge \cite{Has-Lusch}.
These conjectures hold to a good precision
for the lowest $n$ and $| \nu |$,
if $L$ exceeds a lower limit (a little more than 1 fm) \cite{RMT}. 
Then the fit determines the scalar condensate $\Sigma$.
Fig.\ \ref{RMTaxcorfig} (left) shows our results for $\la z_{1} \ra$ 
with $D_{\rm ov-W}$ and $D_{\rm ov-HF}$,
in $V=(1.48~{\rm fm})^{3} \times 2.96~{\rm fm}$.
The RMT conjectures work best in 
the sectors $| \nu | =0,1,2$, and they provide precise
values for $\Sigma$; for $D_{\rm ov-HF}$ we obtain 
$\Sigma = {\bf (268(2) ~ {\rm\bf MeV} )^{3}}$.

\FIGURE{
  \centering
  \includegraphics[angle=270,width=.49\linewidth]{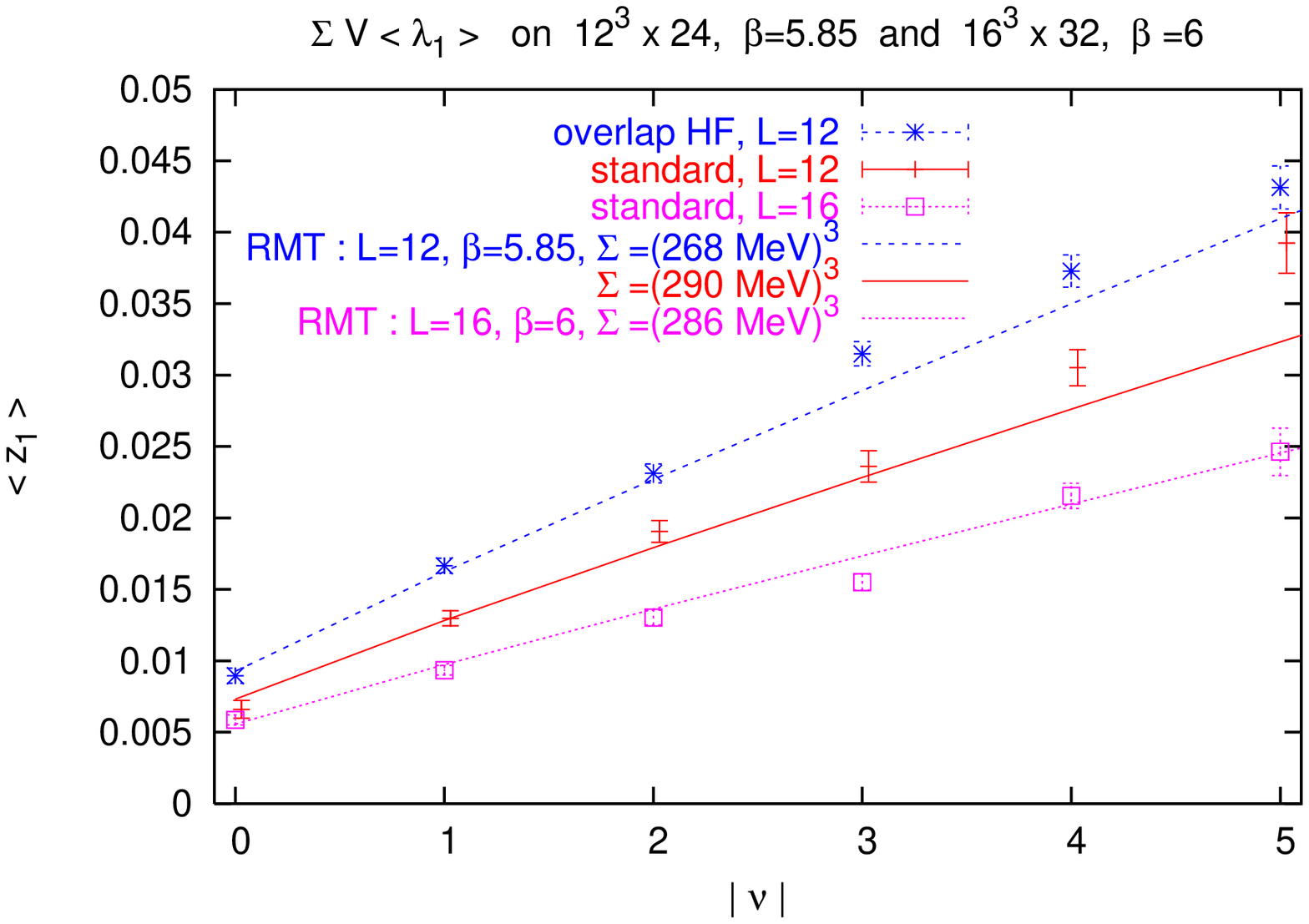}
  \includegraphics[angle=270,width=.49\linewidth]{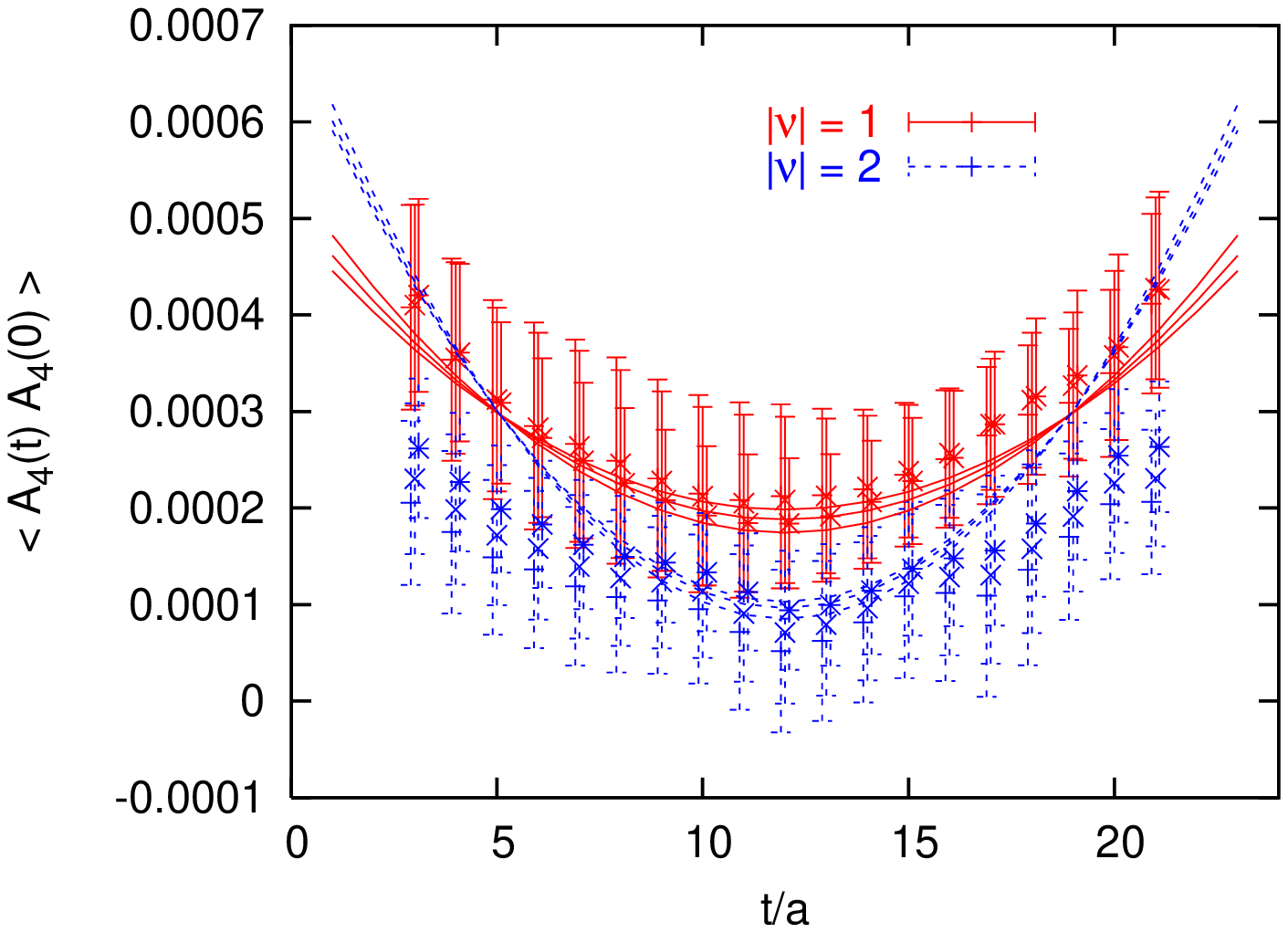}
 \caption{On the left: the lowest Dirac eigenvalue 
(re-scaled with $\Sigma V$) compared the
the RMT predictions at the optimal value of $\Sigma$.
On the right: preliminary results for the axial correlator
at $am_{q} = 0.001$, $0.003$ and $0.005$
in the sectors $| \nu |
=1$ and $2$, and fits to the formulae of quenched $\chi$PT.}
\label{RMTaxcorfig}
}

In quenched $\chi$PT, the axial vector correlator 
depends in leading order only on the LEC
$\Sigma $ and $F_{\pi}$ \cite{qXPT2}. 
The prediction for $\la A_{4} (t) A_{4} (0)\ra$
(with $A_{\mu}(t):= \sum_{\vec x} \bar \psi(\vec x, t) \gamma_{5} 
\gamma_{\mu} \psi (\vec x , t)$) is a parabola with a minimum at $t = T/2$, 
where $F_{\pi}^{2} /T$ enters as an additive constant. In a previous study
we observed that $L$ should again be above $1 ~{\rm fm}$, and that the
history in the sector $\nu =0$ may be plagued by spikes \cite{AA}; 
``Low Mode Averaging'' was then invented as a remedy
\cite{LMA}. In the non-trivial sectors $\Sigma$ can hardly be determined, 
but $F_{\pi}$ can be evaluated. 
In Fig.\ \ref{RMTaxcorfig} (right) we show 
preliminary results from three $m_{q}$ values in the $\eps$-regime, in the
sector $| \nu | = 1$ and $2$, with 10 propagators in each case.
Inserting the RMT result for $\Sigma$, a global fit yields 
bare $F_{\pi}^{(0)} = (96 \pm 10) ~ {\rm MeV}$, which is renorma\-lised 
with $Z_{A}=1.17$ to ${\bf{F_{\pi} = (105 \pm 13)}} ~ {\rm{\bf{ MeV}}}$, 
in agreement with other quenched results \cite{LMA,zeromode,Jap}.

\FIGURE{
  \centering
  \includegraphics[angle=270,width=.48\linewidth]{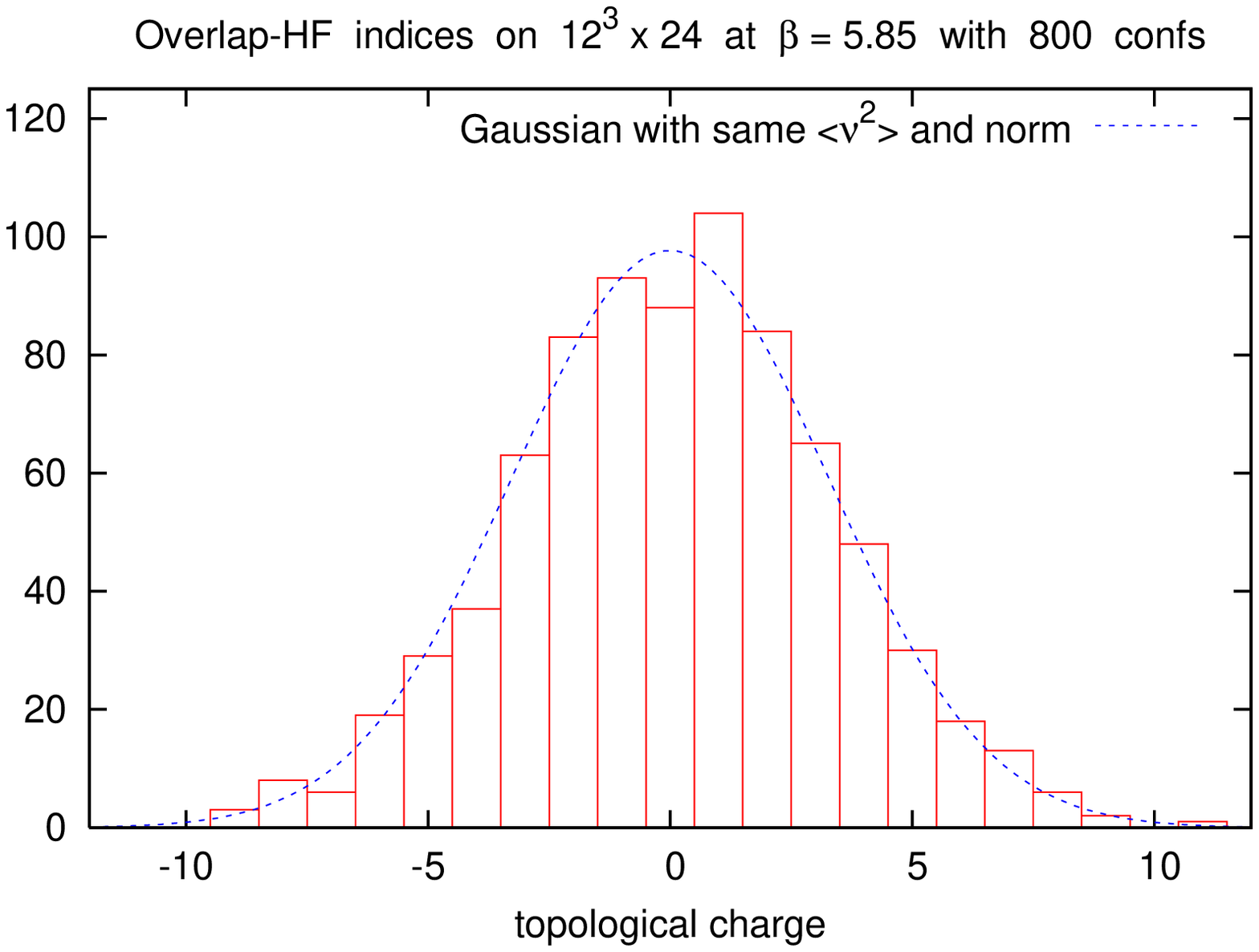}
  \includegraphics[angle=270,width=.48\linewidth]{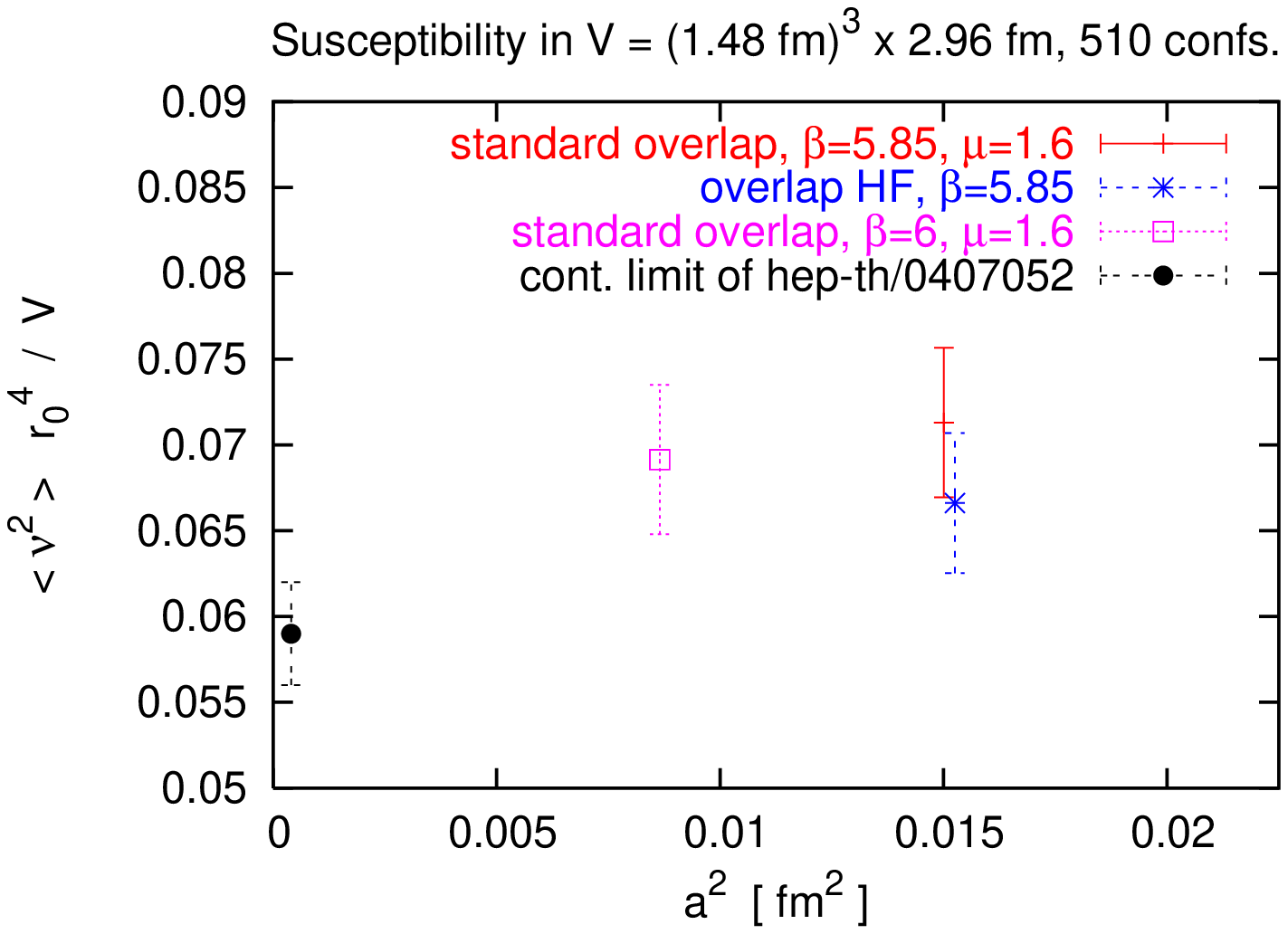}
 \caption{Histogram of overlap HF indices, and
results for the topological susceptibility.}
\label{topofig}
}

For 800 overlap HF indices we obtained the histogram in Fig.\
\ref{topofig} (left). The double peak reminds us of the
question if parity could be broken \cite{parity},
but the current statistics is of course not conclusive for this point.
Fig.\ \ref{topofig} (right) shows our results for the topological 
susceptibility, compared to the continuum extrapolation of Ref.\ \cite{DGP},
which used $D_{\rm ov-W}$.  There is no contradiction, although our 
susceptibilities are somewhat larger. The $D_{\rm ov-HF}$ result
is closer to the value of Ref.\ \cite{DGP} (for exactly the same configurations
at $\beta = 5.85$,
with $\la | \nu _{\rm ov-W} - \nu_{\rm ov-HF} | \ra \approx 0.8$) \cite{Rum}.

\section{Conclusions}\label{section4}

The overlap HF operator provides better locality, and therefore chiral
fermion on coarser lattices than the standard overlap operator.
In the $p$-regime we gave results for $m_{\pi}$, $m_{\rho}$ and 
$F_{\pi}$. Compared to the standard overlap fermion, $m_{\rm PCAC}$ is closer to
the bare quark mass $m_{q}$, hence $Z_{A}$ is much closer to $1$.
In the $\eps$-regime the confrontation with RMT yields a precise value
for $\Sigma$, and from the axial correlation we extracted a preliminary
result for $F_{\pi}$, which agrees (within the errors) with the
chiral extrapolation of the direct measurement in the $p$-regime.
In the $\eps$-regime one may consider as an alternative
solely the $0$-mode contributions to the mesonic correlators 
\cite{zeromode,Stani,Rum}. Finally we add
that a topology conserving gauge action could be helpful in that regime
\cite{topogauge}.

\vspace*{3mm}

{\small\it We are indebted to M.\ Papinutto and C.\ Urbach for 
valuable numerical tools. We also thank them and
B.\ Alles, L.\ Del Debbio,
S. D\"{u}rr, H.\ Fukaya, K.-I.\ Nagai, K.\ Ogawa, L.\ Scorzato, 
A.\ Shindler, H.\ St\"{u}ben and U.\ Wenger for useful communications. 
This work was supported by the Deutsche Forschungsgemeinschaft 
through SFB/TR9-03. 
The computations were performed on the IBM p690 clusters of the HLRN
``Norddeutscher Verbund f\"ur Hoch- und H\"ochstleistungsrechnen'' (HLRN) 
and at NIC, Forschungszentrum J\"{u}lich.}

\end{document}